# Study of the Decay $\Lambda_c^+ \to \Lambda \ell^+ \nu_l$

T. Bergfeld,[1] B.I. Eisenstein,[1] G. Gollin,[1] B. Ong,[1] M. Palmer,[1] M. Selen,[1] J. J. Thaler,[1] A.J. Sadoff,[2] R. Ammar,[3] S. Ball,[3] P. Baringer,[3] A. Bean,[3] D. Besson,[3] D. Coppage,[3] N. Copty,[3] R. Davis,[3] N. Hancock,[3] M. Kelly,[3] N. Kwak,[3] H. Lam,[3] Y. Kubota,[4] M. Lattery,[4] J.K. Nelson,[4] S. Patton,[4] D. Perticone,[4] R. Poling,[4] V. Savinov,[4] S. Schrenk,[4] R. Wang,[4] M.S. Alam,[5] I.J. Kim,[5] B. Nemati,[5] J.J. O'Neill,[5] H. Severini,[5] C.R. Sun,[5] M.M. Zoeller,[5] G. Crawford,[6] C. M. Daubenmier,[6] R. Fulton,[6] D. Fujino,[6] K.K. Gan,[6] K. Honscheid,[6] H. Kagan,[6] R. Kass,[6] J. Lee,[6] R. Malchow,[6] F. Morrow,[6] Y. Skovpen,[6*] M. Sung,[6] C. White,[6] F. Butler,[7] X. Fu,[7] G. Kalbfleisch,[7] W.R. Ross,[7] P. Skubic,[7] J. Snow,[7] P.L. Wang,[7] M. Wood,[7] D.N. Brown,[8] J.Fast ,[8] R.L. McIlwain,[8] T. Miao,[8] D.H. Miller,[8] M. Modesitt,[8] D. Payne,[8] E.I. Shibata,[8] I.P.J. Shipsey,[8] P.N. Wang,[8] M. Battle,[9] J. Ernst,[9] Y. Kwon,[9] S. Roberts,[9] E.H. Thorndike,[9] C.H. Wang,[9] J. Dominick,[10] M. Lambrecht,[10] S. Sanghera,[10] V. Shelkov,[10] T. Skwarnicki,[10] R. Stroynowski,[10] I. Volobouev,[10] G. Wei,[10] P. Zadorozhny,[10] M. Artuso,[11] M. Goldberg,[11] D. He,[11] N. Horwitz,[11] R. Kennett,[11] R. Mountain,[11] G.C. Moneti,[11] F. Muheim,[11] Y. Mukhin,[11] S. Playfer,[11] Y. Rozen,[11] S. Stone,[11] M. Thulasidas,[11] G. Vasseur,[11] G. Zhu,[11] J. Bartelt,[12] S.E. Csorna,[12] Z. Egyed,[12] V. Jain,[12] K. Kinoshita,[13] K.W. Edwards,[14] M. Ogg,[14] D.I. Britton,[15] E.R.F. Hyatt,[15] D.B. MacFarlane,[15] P.M. Patel,[15] D.S. Akerib,[16] B. Barish,[16] M. Chadha,[16] S. Chan,[16] D.F. Cowen,[16] G. Eigen,[16] J.S. Miller,[16] C. O'Grady,[16] J. Urheim,[16] A.J. Weinstein,[16] D. Acosta,[17] M. Athanas,[17] G. Masek,[17] H.P. Paar,[17] J. Gronberg,[18] R. Kutschke,[18] S. Menary,[18] R.J. Morrison,[18] S. Nakanishi,[18] H.N. Nelson,[18] T.K. Nelson,[18] C. Qiao,[18] J.D. Richman,[18] A. Ryd,[18] H. Tajima,[18] D. Schmidt,[18] D. Sperka,[18] M.S. Witherell,[18] M. Procario,[19] R. Balest,[20] K. Cho,[20] M. Daoudi,[20] W.T. Ford,[20] D.R. Johnson,[20] K. Lingel,[20] M. Lohner,[20] P. Rankin,[20] J.G. Smith,[20] J.P. Alexander,[21] C. Bebek,[21] K. Berkelman,[21] K. Bloom,[21] T.E. Browder,[21] D.G. Cassel,[21] H.A. Cho,[21] D.M. Coffman,[21] P.S. Drell,[21] R. Ehrlich,[21] M. Garcia-Sciveres,[21] B. Geiser,[21] B. Gittelman,[21] S.W. Gray,[21] D.L. Hartill,[21] B.K. Heltsley,[21] C.D. Jones,[21] S.L. Jones,[21] J. Kandaswamy,[21] N. Katayama,[21] P.C. Kim,[21] D.L. Kreinick,[21] G.S. Ludwig,[21] J. Masui,[21] J. Mevissen,[21] N.B. Mistry,[21] C.R. Ng,[21] E. Nordberg,[21] J.R. Patterson,[21] D. Peterson,[21] D. Riley,[21] S. Salman,[21] M. Sapper,[21] F. Würthwein,[21] P. Avery,[22] A. Freyberger,[22] J. Rodriguez,[22] R. Stephens,[22] S. Yang,[22] J. Yelton,[22] D. Cinabro,[23] S. Henderson,[23] T. Liu,[23] M. Saulnier,[23] R. Wilson,[23] and H. Yamamoto[23]

(CLEO Collaboration)

[1]University of Illinois, Champaign-Urbana, Illinois, 61801
[2]Ithaca College, Ithaca, New York 14850
[3]University of Kansas, Lawrence, Kansas 66045
[4]University of Minnesota, Minneapolis, Minnesota 55455
[5]State University of New York at Albany, Albany, New York 12222
[6]Ohio State University, Columbus, Ohio, 43210
[7]University of Oklahoma, Norman, Oklahoma 73019
[8]Purdue University, West Lafayette, Indiana 47907
[9]University of Rochester, Rochester, New York 14627
[10]Southern Methodist University, Dallas, Texas 75275
[11]Syracuse University, Syracuse, New York 13244
[12]Vanderbilt University, Nashville, Tennessee 37235
[13]Virginia Polytechnic Institute and State University, Blacksburg, Virginia, 24061
[14]Carleton University, Ottawa, Ontario K1S 5B6 and the Institute of Particle Physics, Canada
[15]McGill University, Montréal, Québec H3A 2T8 and the Institute of Particle Physics, Canada
[16]California Institute of Technology, Pasadena, California 91125
[17]University of California, San Diego, La Jolla, California 92093
[18]University of California, Santa Barbara, California 93106
[19]Carnegie-Mellon University, Pittsburgh, Pennsylvania 15213
[20]University of Colorado, Boulder, Colorado 80309-0390
[21]Cornell University, Ithaca, New York 14853
[22]University of Florida, Gainesville, Florida 32611
[23]Harvard University, Cambridge, Massachusetts 02138

## Abstract

Using the CLEO II detector at CESR we observe 500 $\Lambda\ell^+$ pairs consistent with the semileptonic decay $\Lambda_c^+ \to \Lambda\ell^+\nu_l$. We measure $\sigma(e^+e^- \to \Lambda_c^+ X)\cdot B(\Lambda_c^+ \to \Lambda\ell^+\nu_l) = 4.77 \pm 0.25 \pm 0.66$pb. Combining with the charm semileptonic width and the lifetime of the $\Lambda_c$ we also obtain $B(\Lambda_c^+ \to pK^-\pi^+)$. We find no evidence for $\Lambda\ell^+\nu_l$ final states in which there are additional $\Lambda_c^+$ decay products. We measure the decay asymmetry parameter of $\Lambda_c^+ \to \Lambda e^+\nu_e$ to be $\alpha_{\Lambda_c} = -0.89^{+0.17}_{-0.11}{}^{+0.09}_{-0.05}$.

*Permanent address: INP, Novosibirsk, Russia

Charm semileptonic decays allow an absolute measurement of the heavy quark decay form factors because the Cabibbo-Kobayashi-Maskawa (CKM) matrix element $V_{cs}$ is known from unitarity. In the spectator model, $\Lambda$-type baryon semileptonic decays are simpler than charm meson semileptonic decays. In the meson case, either a $K^*$ or a $K$ is produced depending on whether the quark spins are parallel or anti-parallel. Conversely, the $\Lambda_c$ spin is carried by the $c$ quark, since the $(ud)$ combination is spin and isospin zero, so that only a $\Lambda$ is formed when the $c$ quark decays. The simplicity of $\Lambda$-type semileptonic decays allows for more reliable predictions, within the framework of Heavy Quark Effective Theory (HQET) [1], concerning heavy quark to light quark transitions [2] [3] than is the case for mesons. In addition, as the $\Lambda_c$ semileptonic decay branching ratio can be reliably predicted from the known inclusive charm semileptonic width and the $\Lambda_c$ lifetime, a measurement of $\sigma(e^+e^- \to \Lambda_c^+ X) \cdot B(\Lambda_c^+ \to \Lambda \ell^+ \nu_l)$ provides an absolute normalization for all $\Lambda_c$ decays.

In this paper we report a measurement of $\sigma(e^+e^- \to \Lambda_c^+ X) \cdot B(\Lambda_c^+ \to \Lambda \ell^+ \nu_l)$. From this we obtain $B(\Lambda_c^+ \to pK^-\pi^+)$. Finally, we use the decay $\Lambda \to p\pi$ as a polarization analyzer to measure the decay asymmetry in $\Lambda_c^+ \to \Lambda e^+ \nu_e$.

The data sample used in this study contains some 2 million $e^+e^- \to c\,\overline{c}$ events collected with the CLEO II detector [4] at the Cornell Electron Storage Ring (CESR). The integrated luminosity consists of 1.1 $fb^{-1}$ taken at the $\Upsilon(4S)$ resonance and 0.5 $fb^{-1}$ taken just below the $B\overline{B}$ threshold.

We search for the decay $\Lambda_c^+ \to \Lambda \ell^+ \nu_l$ by detecting a $\Lambda \ell^+$ (right sign) pair with invariant mass in the range $m_\Lambda < m_{\Lambda\ell} < m_{\Lambda_c}$ [5]. A background source of right sign $\Lambda\ell$ pairs is the semileptonic decay of the heavier charm baryons $\Xi_c^0$, $\Xi_c^+$, and $\Omega_c^0$ . Secondary background sources that can produce both right and wrong sign $\Lambda\ell$ pairs are the continuum production of $\Lambda$'s not associated with charm baryons, denoted by "$c\overline{c}$", and B decays at the $\Upsilon(4S)$, denoted by "$B\overline{B}$". The majority of $\Lambda\ell^-$ (wrong sign) pairs will be produced either from $e^+e^- \to \Lambda_c\overline{\Lambda}_c$ where one $\Lambda_c$ decays to a $\Lambda$ and the other decays semileptonically or from $e^+e^- \to \Lambda_c \overline{M}_c \overline{N}$ where $\overline{M_c}$ is a charm meson that decays semileptonically. In both of these cases the $\Lambda\ell$ invariant mass will often satisfy $m_{\Lambda\ell} > m_{\Lambda_c}$. The wrong sign sample is used to normalize our Monte Carlo (MC) simulation of secondary background sources.

We search each event for a $\Lambda\ell$ combination. All tracks are required to come from the region of the event vertex. Electrons are identified using a likelihood function which incorporates information from the calorimeter and dE/dx systems, and muons are identified by their ability to penetrate an iron absorber and reach detection planes at a depth of at least five nuclear absorption lengths [4]. The minimum allowed momentum is 0.7 GeV/c for electrons and 1.4 GeV/c for muons. Leptons are required to have been detected in the so-called barrel region, i.e. $|\cos\theta| < 0.71$, where $\theta$ is the angle of the lepton momentum with the beam line. To reduce the background from $B$ decays, we require $R_2 = H_2/H_0 > 0.2$ where $H_i$ are Fox-Wolfram event shape variables [6].

The $\Lambda$ is reconstructed through its decay to $p\pi$. We require the point of intersection of the two charged tracks, measured in the $r-\phi$ plane, to be greater than 0.5 cm away from the primary vertex. In addition, we require the sum of the $p$ and $\pi$ momentum vectors to extrapolate back to the beamline. The $dE/dx$ measurement of the proton is required to be consistent with the expected value. We reject combinations which satisfy interpretation as a $K^0_s$. Finally, we require the momentum of the $p\pi$ pair to be greater than 0.8 GeV/c in order to reduce background from secondary sources.

These $\Lambda$ candidates are then combined with leptons, and the sum of $\Lambda$ and $\ell$ momentum, $p_{\Lambda\ell}$, is required to be greater than 1.4 GeV/c. This cut reduces our dependence on the shape of the $\Lambda_c$ fragmentation function at low momentum, which is poorly known. The invariant mass of the $\Lambda\ell$ pair is required to satisfy $1.3 < m_{\Lambda\ell} < 2.3$ GeV/$c^2$. To determine the number of events in the signal region we fit the $p\pi$ invariant mass distribution with a function consisting of a Gaussian, whose width is determined by a MC simulation, and a polynomial background. The fits for the electron and muon samples, in both right sign and wrong sign combinations, are shown in Figure 1 and the results of the fits are given in Table I.

In addition to background from the sources outlined above, $\Lambda\ell$ pairs are also produced by combining real leptons with fake $\Lambda$'s and fake leptons with real $\Lambda$'s. The first of these is already taken into account by fitting to the $\Lambda$ mass to determine the yield. Care must be taken in calculating the wrong sign fake background because antiprotons are particularly likely to be misidentified as electrons and because they are produced copiously in association with $\Lambda$'s due to baryon conservation. In order to take into account the different fake rates of protons and antiprotons compared to kaons and pions, we multiply the number of tracks, which are not positively identified as lepton tracks, by the the fake probabilities weighted by the particle population given by the LUND model [7] for continuum events containing a $\Lambda$. We fit the $p\pi$ invariant mass distribution to determine the number of fake lepton with real $\Lambda$ combinations. The results are given in Table I for both right and wrong sign combinations.

The MC was used to estimate the number of real, right and wrong sign $\Lambda\ell$ combinations produced by secondary sources satisfying our selection criteria. To test our MC, we predict the number of wrong sign events both in the signal region and for $m_{\Lambda\ell} > 2.3$ GeV/c$^2$. When the number of wrong sign, fake lepton events is subtracted from the number of wrong sign events, the MC prediction is in good agreement with the data over the whole mass range. We therefore have confidence in the MC predictions for the right sign secondary sources. The predicted number of events for the electron channel are $10.2 \pm 10.2$ for $B\overline{B}$ and $21.6 \pm 21.6$ for $c\overline{c}$. The large errors account for model uncertainty.

In Figure 2, we show the right sign $m_{\Lambda\ell}$ distribution after $\Lambda$ sideband subtraction [8]. Agreement between the data and MC is good. However, there is a slight excess of events in the data at low mass. There would be a low mass excess if there are contributions other than $\Lambda_c^+ \to \Lambda\ell^+\nu_l$. Such decays would have a lower efficiency, and would lead to an underestimation of $\sigma \cdot B$. We have made an extensive search for possible decays of the type $\Lambda_c^+ \to \Lambda X e^+ \nu_e$, where $X$ represents additional decay products. We only search for the lowest lying excited baryons because our efficiency for reconstructing semileptonic decays involving heavier baryons is very small. The decay modes we examine are:

- $\Lambda_c^+ \to \Lambda^* e^+ \nu_e, \Lambda^* \to \Sigma^0 \pi^0$ with $\Sigma^0 \to \Lambda\gamma$,

  where $\Lambda^* = \Lambda(1405), \Lambda(1520), \Lambda(1600), \Lambda(1670), \Lambda(1690)$

- $\Lambda_c^+ \to \Sigma^* X e^+ \nu_e, \Sigma^* \to \Lambda\pi$, where $\Sigma^* = \Sigma^+(1385), \Sigma^+(1660), \Sigma^0(1385), \Sigma^0(1660)$

- $\Lambda_c^+ \to \Sigma^0 X e^+ \nu_e, \Sigma^0 \to \Lambda\gamma$ and $\Lambda_c^+ \to \Lambda(\pi\pi)^0 e^+ \nu_e$

In addition to the selection criteria previously described, we select photon candidates from showers in the calorimeter that have a minimum energy of 40 MeV, are not matched to a charged-particle track from the drift chamber, and have a lateral energy distribution consistent with that expected for photons. Neutral pion candidates are selected from pairs of photons with at least one photon in the barrel portion of the calorimeter. We also require the two photon mass to be within 3 standard deviations of the known pion mass. All selected candidates are kinematically fitted to the $\pi^0$ mass.

We observe no events of the type $\Lambda_c^+ \to \Lambda X \ell^+ \nu_l$. The results of all the searches are given in Table II. Combining the results, we find the number of events attributable to $\Lambda_c^+ \to \Lambda X \ell^+ \nu_l$ is less than 15% of the signal events at 90% confidence level.

Our MC simulation found that, for the selection criteria used in this analysis, the efficiency of reconstructing the $\Xi_c^0$ and $\Xi_c^+$ modes is about 35% of that for $\Lambda_c^+ \to \Lambda \ell^+ \nu_l$. Using the relative efficiency and measurement of $\sigma \cdot B$ for the semileptonic decay of $\Xi_c^0$ and the measured ratio of lifetimes $\tau(\Xi_c^+)/\tau(\Xi_c^0)$ [9], we obtain the background given in Table I. (We do not include a contribution from $\Omega_c^0$ semileptonic decays because they have never been observed, and the production cross section for the $\Omega_c^0$ is expected to be small.)

The efficiencies given in Table I are obtained by MC simulation and include $B(\Lambda \to p\pi)$. We obtain the yield given in Table I by subtracting the fake contribution and the $\Xi_c^0$ and $\Xi_c^+$ feedthrough background from the number of right sign events. The statistical error in the yield is calculated from the number of right sign events. The efficiency-corrected yield and integrated luminosity $\mathcal{L}$ are used to obtain the $\sigma \cdot$ B given in Table I.

As the $B\overline{B}$ and $c\bar{c}$ backgrounds are small and model dependent, we choose not to subtract them but instead to incorporate them into the systematic error. The largest experimental source of the systematic error in $\sigma \cdot$ B is the uncertainty in the fake rates (7%). We have investigated model dependence by varying the fragmentation function (5%) and by taking the difference (6%) in calculated efficiencies from the HQET Körner–Krämer (KK) model [3] (which is the efficiency used for the result) and a semileptonic decay model producing $\Lambda$'s with no net polarization (model A). An additional systematic error of 7% comes from the possible contribution of $\Lambda_c^+$ decays other than $\Lambda_c^+ \to \Lambda \ell^+ \nu_l$.

Our results are :

$$\sigma \cdot B(\Lambda_c^+ \to \Lambda e^+ \nu_e) = (4.87 \pm 0.28 \pm 0.69) \text{ pb}$$

$$\sigma \cdot B(\Lambda_c^+ \to \Lambda \mu^+ \nu_\mu) = (4.43 \pm 0.51 \pm 0.64) \text{ pb.}$$

As the two results are statistically independent, we form the weighted average:

$$\sigma \cdot \text{B} = (4.77 \pm 0.25 \pm 0.66) \text{ pb.}$$

Our result, which is in agreement with a measurement from ARGUS [10], is the most precise measurement of this quantity to date.

At present, all $\Lambda_c$ branching ratios are normalized to $B(\Lambda_c \to pK\pi)$, $B_{pK\pi}$, the measurements of which are model dependent and vary by a factor of two [11]. We can determine $B_{pK\pi}$ from our measurement of $\sigma \cdot$ B. As semileptonic decay proceeds only via the emission of a virtual $W$ from the heavy quark, the semileptonic width of all charmed hadrons should be equal. This is well established for the $D^+$ and $D^0$ [12] where: $\Gamma_{sl}^{D^+} = (1.61 \pm 0.18) \times 10^{11} s^{-1}$ and $\Gamma_{sl}^{D^0} = (1.83 \pm 0.29) \times 10^{11} s^{-1}$. This equality should also hold for the $\Lambda_c$, assuming no new process gives rise to leptons in baryonic weak decay. Therefore, we can predict the inclusive semileptonic branching ratio, $B_{\ell X}$, to be:

$$B_{\ell X} = \mathrm{B}(\Lambda_c^+ \to e^+ X) =< \Gamma_{sl} > \tau_{\Lambda_c} = (3.4 \pm 0.4)\%$$

where we have used the weighted average of the semileptonic widths for $D^0$ and $D^+$ and the average lifetime of $\Lambda_c$ [13]. There exists no theoretical relationship between $B(\Lambda_c^+ \to \Lambda X \ell^+ \nu_l)$ and $B_{\ell X}$. Also, it is probable that there are semileptonic decays of the $\Lambda_c$ which do not include a $\Lambda$, such as $\Lambda_c^+ \to \Sigma^+ \pi^- \ell^+ \nu_l$, $\Lambda_c^+ \to pK^- \ell^+ \nu_l$, as well as Cabibbo suppressed decays like $\Lambda_c^+ \to n \ell^+ \nu_l$. We therefore expect $f \equiv \frac{B(\Lambda_c^+ \to \Lambda \ell^+ \nu_l)}{B_{\ell X}} \leq 1$.

We use the experimentally determined $\Lambda_c$ fragmentation function in the MC simulation. For $x_p > 0.5$ [14], the weighted average of the muon and electron samples is $\sigma \cdot \mathrm{B} = (3.38 \pm 0.18 \pm 0.47)$ pb. Combining this result with the CLEO measurement of $\sigma(e^+e^- \to \Lambda_c^+ X) \cdot B(\Lambda_c^+ \to pK^-\pi^+)$ [15] gives $R \equiv \frac{\sigma \cdot B_{pK\pi}}{\sigma \cdot \mathrm{B}} = 1.93 \pm 0.10 \pm 0.33$, which is independent of the details of $\Lambda_c$ production. Therefore:

$$B_{pK\pi} = f \; R \; B_{\ell X} = f \; (6.67 \pm 0.35 \pm 1.35)\%$$

This result does not exclude any of the previous measurements of $B_{pK\pi}$ but places a reliable upper bound on this important but poorly determined quantity and thereby provides an absolute scale for all $\Lambda_c$ branching ratios.

In the spectator model the quantity that corresponds to $f$ in the charmed meson sector is $\Gamma(D \to (K^* + K)\ell\nu_l)/\ \Gamma(D \to \ell X) = 0.89 \pm 0.12$ [16]. Using this value, our result is compatible with the CLEO and ARGUS determination of $B_{pK\pi}$ from B decay at the $\Upsilon(4S)$ [11].

In semileptonic baryon decay it is possible to construct six invariant hadronic amplitudes, each of which is a function of $q^2$ (the mass squared of the virtual $W$), from the spins and momenta of the initial and final state baryons. In the framework of HQET, the heavy flavor and spin symmetries imply relations among these amplitudes resulting in only one universal form factor when the final state also includes a heavy quark. If the final state contains only light quarks, as in $\Lambda_c$ decay, then it can be shown that two independent form factors, $f_1$ and $f_2$, are required. For any heavy $\Lambda$-type baryon decays, HQET predicts $G_A = -G_V$ [2] [3], where $G_V$ is the vector coupling and $G_A$ is the axial vector coupling of the hadronic current. The physical consequence is that the daughter baryon will be emitted with 100% negative polarization at $q^2 = 0$. The physical observable is the angle, $\Theta_\Lambda$, between the momentum vector of the proton (or pion) in the $\Lambda$ rest frame and the $\Lambda$ momentum in the $\Lambda_c$ rest frame. One finds:

$$\frac{d\Gamma}{dq^2 dcos\Theta_\Lambda} \propto 1 + \alpha_{\Lambda_c}\alpha_\Lambda cos\Theta_\Lambda$$

where $\alpha_{\Lambda_c}$ ($\alpha_\Lambda$) is the decay asymmetry parameter of the $\Lambda_c$ ($\Lambda$) and $\alpha_\Lambda = 0.64$ [12].

The $q^2$-dependence of $\alpha_{\Lambda_c}$ depends on the details of the form factor structure. If the two form factors are assumed to have the same $q^2$ dependence, then $\alpha_{\Lambda_c}$ depends only on the ratio of the form factors, $r = f_2/f_1$, which is expected to be less than unity. Over most of the $q^2$ range $\alpha_{\Lambda_c}$ is less than –0.5 for reasonable values of $r$ and is exactly –1 at $q^2 = 0$ for all $r$ [3]. Including terms of order $(1/m_c)$ in the HQET expansion leads to a small departure from –1 at $q^2 = 0$ [17].

We have shown that the $\Lambda\ell^+$ signal events are predominantly $\Lambda_c^+ \to \Lambda\ell^+\nu_l$. We use only $\Lambda e$ events to measure the asymmetry as the muon sample is very small.

Measuring $\cos\Theta_\Lambda$ requires knowledge of the $\Lambda_c$ momentum but this is unknown because the neutrino is unobserved. We estimate the direction of the $\Lambda_c$ from the thrust axis of the event. The magnitude of the $\Lambda_c$ momentum is then obtained by solving the equation $\vec{P}^2_{\Lambda_c} = (\vec{P}_\Lambda + \vec{P}_e + \vec{P}_{\nu_e})^2$. The two solutions correspond physically to the undetermined longitudinal momentum of the neutrino. If the direction of the $\Lambda_c$ was known exactly, one of the two solutions would have a value equal to the true magnitude of the $\Lambda_c$ momentum. As the thrust axis is only an approximation to the true $\Lambda_c$ direction, neither solution corresponds to the true $\Lambda_c$ momentum. We therefore determine the best solution by MC simulation. We generate events according to the KK model for five values of $r$ over the theoretically expected range $-0.5 < r < 0.5$ (-0.97< $\alpha_{\Lambda_c}$ <-0.61). The results of our simulation are:

1. In about half of the events, the thrust axis is not well aligned with the true $\Lambda_c$ direction resulting in two non-physical solutions. In this case the direction of the thrust vector is varied until a physical solution results.

2. There is one physical solution in about 10% of the cases.

3. When there are two physical solutions, the higher (lower) momentum solution systematically over-estimates (under-estimates) the $\Lambda_c$ momentum by similar amounts. In consequence, a measurement of $\cos\Theta_\Lambda$ which chooses the higher (lower) solution systematically under-estimates (over-estimates) the $\Lambda$ polarization. We find that the best estimate of the $\Lambda_c$ momentum is obtained from the weighted average of the two solutions, where the weights are given by the measured $\Lambda_c$ fragmentation function [15].

Using this method for estimating the $\Lambda_c$ momentum, the resulting $\alpha_{\Lambda_c}$ is consistent with the input value. The resolution in $\cos\Theta_\Lambda$ is 0.2, independent of the original number of the solutions for the $\Lambda_c$ momentum, so we separate the data into four $\cos\Theta_\Lambda$ bins.

We calculate the $\Lambda_c$ momentum for the data exactly as for the MC. We determine the yield in each $\cos\Theta_\Lambda$ bin by fitting the $p\pi$ invariant mass spectra shown in Figure 3. We compute the fake lepton background as a function of $\cos\Theta_\Lambda$ using the same procedure described above and subtract the fake background from each bin.

The efficiency as a function of $\cos\Theta_\Lambda$ is flat except for a small decrease towards $\cos\Theta_\Lambda = +1$. This bin corresponds to a backward soft pion which is difficult to detect in the drift chamber. We use the KK model with $r = -0.25$ to calculate the efficiency in each $\cos\Theta_\Lambda$ bin. The efficiency corrected $\cos\Theta_\Lambda$ distribution is shown in Figure 4. We find $\alpha_{\Lambda_c} = -0.89^{+0.17}_{-0.11}\,^{+0.09}_{-0.05}$, where the first error is the error returned from the fit and the second error is the systematic error.

We have examined five sources of systematic error in the determination of $\alpha_{\Lambda_c}$: the effect of using the thrust axis as the $\Lambda_c$ direction, modelling of the detector efficiency as a function of $\cos\Theta_\Lambda$, model dependence, the fitting method, and the presence of $\Xi_c^0$, $\Xi_c^+$, and $\Omega_c^0$ decays or decays other than $\Lambda_c^+ \to \Lambda\ell^+\nu_l$. We discuss each of these sources below.

We do not find any evidence in the MC for a shift between the input and reconstructed values of $\alpha_{\Lambda_c}$ for the KK model using five different values of $r$. Therefore we do not assign a systematic error due to the methods of estimating the $\Lambda_c$ momentum.

To allow for errors in efficiency modelling as a function of $\cos\Theta_\Lambda$, we repeat our analysis by varying the efficiency according to our understanding of the detector performance. We ascribe a systematic error of -0.04 from this source.

To estimate the KK model dependence, we repeat the analysis using the efficiency computed for five values of $r$ satisfying -0.5< $r$ <0.5. The spread in values of $\alpha_{\Lambda_c}$ is described by a systematic error of $(^{-0.03}_{+0.01})$. In order to determine the overall model dependence of our result, we repeat the analysis using model A. We find a result which is larger by 0.17 and take half of this difference as the systematic error arising from this source.

We treat the final two systematic errors simultaneously. To allow for the presence of $\Xi_c^0$, $\Xi_c^+$, and $\Omega_c^0$ decays and $\Lambda_c^+$ decays other than $\Lambda_c^+ \to \Lambda\ell^+\nu_l$ in the asymmetry distribution, we add various polynomial background functions to the fit. The area of the polynomial is constrained to that expected from this work and reference [9]. Following this procedure the result of the fit does not change so we do not ascribe a systematic error to this source. The result varies by 0.01 using different fitting methods. This variation is taken as the systematic error.

Our result is consistent with the HQET KK model. The average value of the efficiency corrected $q^2$ distribution of the data is 0.7 $(\mathrm{GeV}/c)^2$. Since we expect $\alpha_{\Lambda_c}$ to decrease as a function of $q^2$ as $q^2 \to 0$, our result implies that $\alpha_{\Lambda_c}$ is close to $-1$ at $q^2 = 0$ in agreement with the prediction of HQET.

We have measured the cross section times branching ratio for the semileptonic decay modes of the $\Lambda_c$ that include a $\Lambda$. We find $\sigma(e^+e^- \to \Lambda_c^+ X) \cdot B(\Lambda_c^+ \to \Lambda e^+\nu_e) = 4.87 \pm$ $0.28\pm0.69$ pb and $\sigma(e^+e^- \to \Lambda_c^+ X)\cdot B(\Lambda_c^+ \to \Lambda\mu^+\nu_\mu) = 4.43\pm0.51\pm0.64$ pb. The combined result is $\sigma(e^+e^- \to \Lambda_c^+ X) \cdot B(\Lambda_c^+ \to \Lambda\ell^+\nu_l) = 4.77 \pm 0.25 \pm 0.66$ pb. Our result is the most precise measurement of this quantity to date. We find the number of events attributable to decays of the type $\Lambda_c^+ \to \Lambda X\ell^+\nu_l$ is less than 15% of the signal events at 90% confidence level. Combining our result for $\sigma \cdot B$ with the charm semileptonic width and the lifetime of the $\Lambda_c$ gives $B(\Lambda_c^+ \to pK^-\pi^+) = f \cdot (6.67 \pm 0.35 \pm 1.35)\%$, where $f$ represents the unknown fraction of $\Lambda_c^+ \to \Lambda\ell^+\nu_l$ to the total semileptonic rate.

A large negative polarization of the $\Lambda$ in the decay $\Lambda_c^+ \to \Lambda e^+\nu_e$ of $\alpha_{\Lambda_c} = -0.89^{+0.17}_{-0.11}\,^{+0.09}_{-0.05}$ has been observed. This is the first observation of parity violation in a semileptonic charmed baryon decay. The sign of this polarization arises from the fact that the hadronic current is V-A. Our result is consistent with the prediction of the HQET KK model. Assuming factorization, one can relate $\alpha_{\Lambda_c}$ for $\Lambda_c^+ \to \Lambda e^+\nu_e$ at $q^2 = m_\pi^2$ to $\alpha_{\Lambda_c}$ in the decay $\Lambda_c \to \Lambda\pi$ determined by CLEO [18] and ARGUS [19]. Our findings are consistent with the factorization ansatz.

We gratefully acknowledge the effort of the CESR staff in providing us with excellent luminosity and running conditions. This work was supported by the National Science Foundation, the U.S. Dept. of Energy, the Heisenberg Foundation, the SSC Fellowship program of TNRLC, and the A.P. Sloan Foundation.

## TABLES

TABLE I. Signals and backgrounds

| mode | electrons | muons |
| --- | --- | --- |
| $N_{\Lambda\ell^+}$ (right) | $510 \pm 25$ | $110 \pm 11$ |
| fakes (right) | $83 \pm 25$ | $13 \pm 4$ |
| $\Xi_c$ feeddown | $77 \pm 24$ | $19 \pm 6$ |
| corrected yield | $350 \pm 25 \pm 28$ | $78 \pm 11 \pm 7$ |
| efficiency (all $x_p$) (%) | $4.44 \pm 0.07$ | $1.09 \pm 0.04$ |
| $\sigma \cdot B(\Lambda_c \to \Lambda\ell^+\nu_l)$ | $4.87 \pm 0.28 \pm 0.69$pb | $4.43 \pm 0.51 \pm 0.64$pb |
| $N_{\Lambda\ell^-}$ (wrong) | $109 \pm 14$ | $7 \pm 5$ |
| fakes (wrong) | $90 \pm 27$ | $21 \pm 6$ |

TABLE II. Upper limits for decays of the type $\Lambda_c^+ \to \Lambda X \ell \nu_l$

| modes | events | efficiency* | upper limit* at 90% CL |
| --- | --- | --- | --- |
| $\Sigma^0 \to \Lambda\gamma$ | $0.0 \pm 8.9$ | 0.46 | 6.6% |
| $\Sigma^+(1385) \to \Lambda\pi^+$ | $0.0 \pm 0.1$ | 0.7 | 0.09% |
| $\Sigma^+(1660) \to \Lambda\pi^+$ | $0.0 \pm 0.1$ | 0.7 | 0.09% |
| $\Sigma^0(1385) \to \Lambda\pi^0$ | $0.0 \pm 0.1$ | 0.3 | 0.1% |
| $\Sigma^0(1660) \to \Lambda\pi^0$ | $0.0 \pm 0.5$ | 0.3 | 0.5% |
| $(\Lambda\pi^+\pi^-)_{nr}\ell^+\nu_l$ | $3.2 \pm 3.8$ | 0.5 | 3.6% |
| $(\Lambda\pi^+\pi^-)_{1520}\ell^+\nu_l$ | $0.4 \pm 6.8$ | 0.5 | 4.5% |

$*$ The efficiencies and upper limits given are relative to that for $\Lambda_c^+ \to \Lambda\ell\nu_l$.

FIGURES

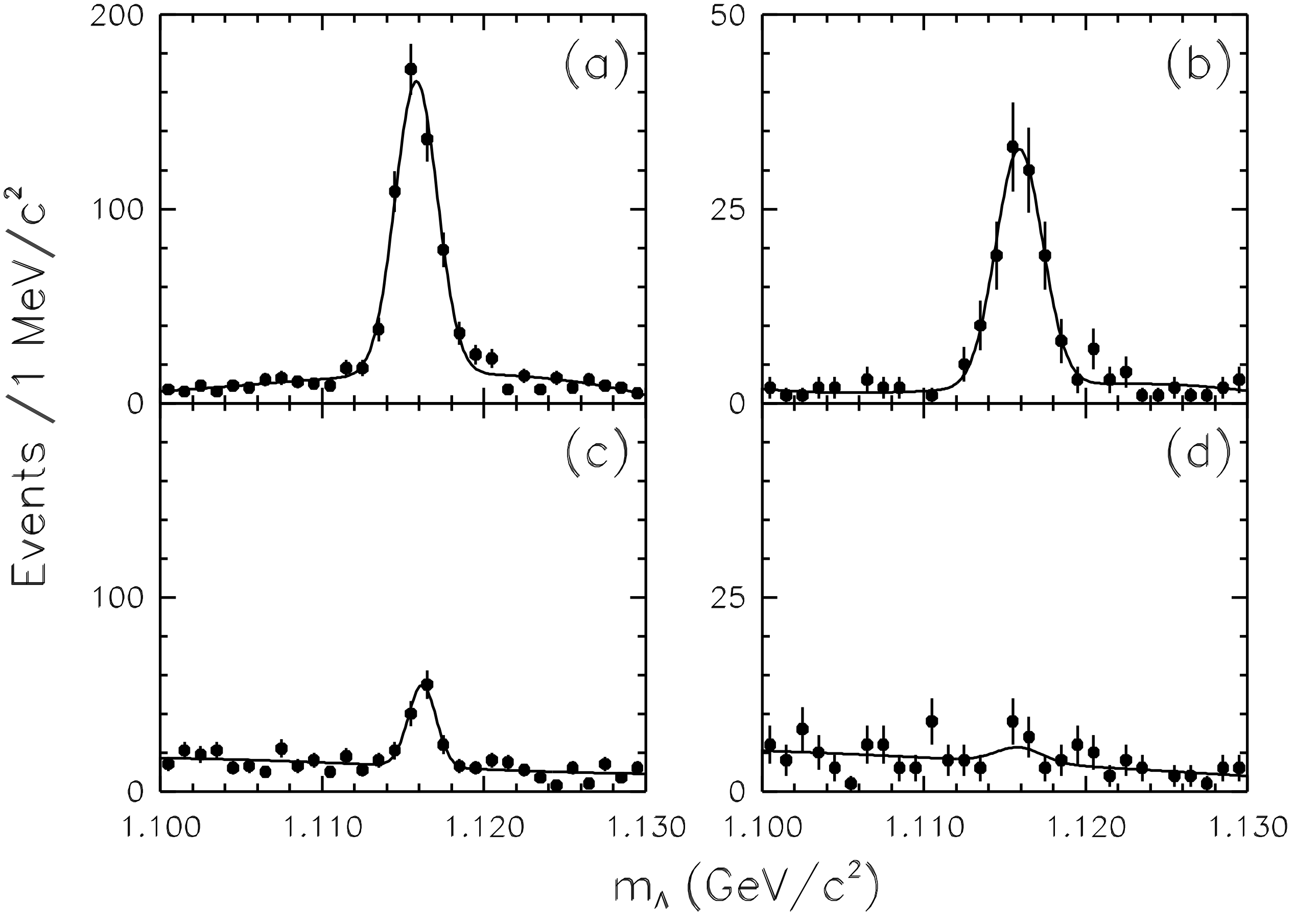

FIG. 1. The $p\pi$ invariant mass for right sign and wrong sign combinations satisfying the cuts described in the text; (a) Right sign electrons, (b) Right sign muons, (c)Wrong sign electrons, (d)Wrong sign muons.

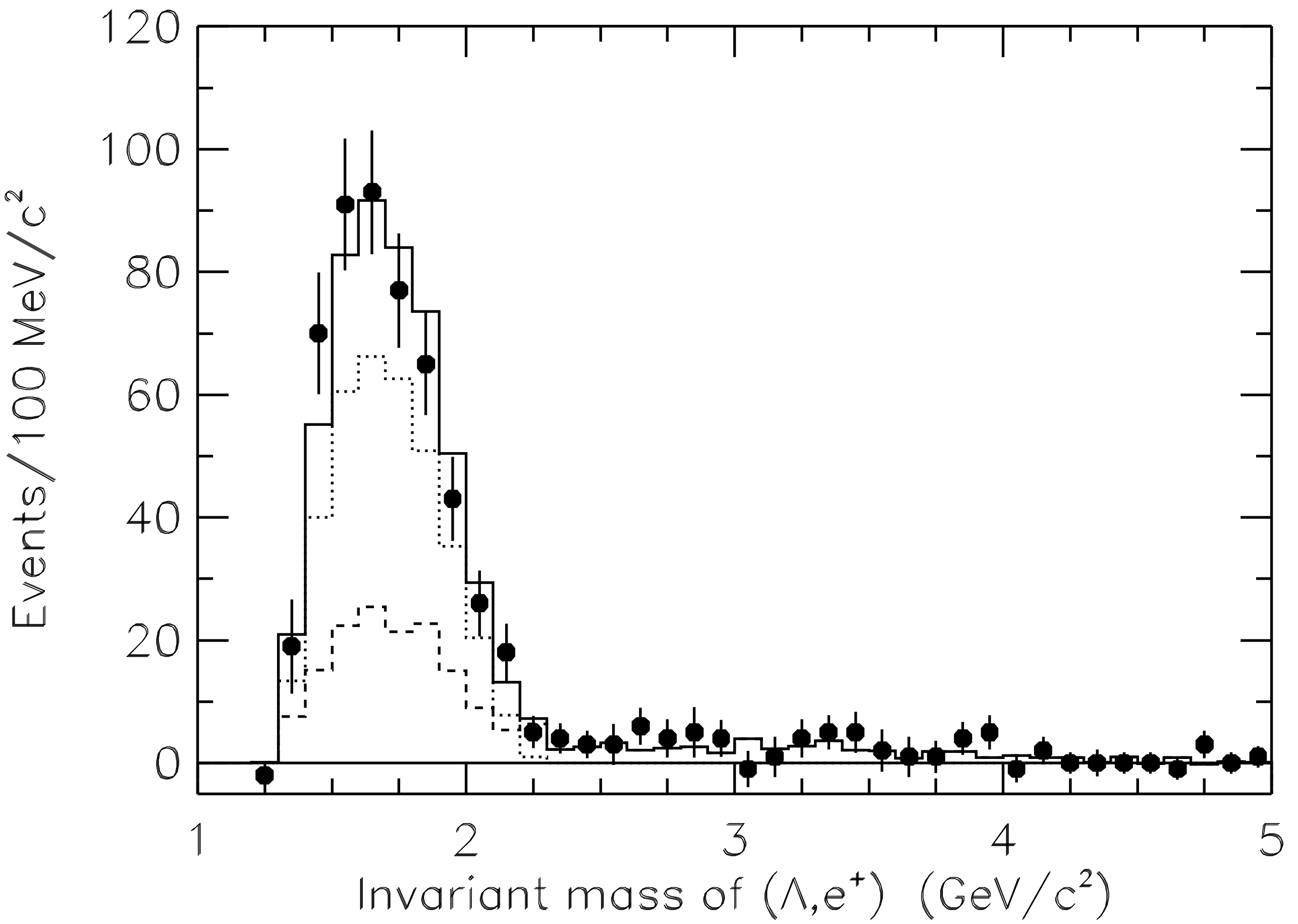

FIG. 2. Invariant $\Lambda e^+$ mass for right sign combinations. The points with error bars are data after subtraction of the contribution of fake lambdas estimated using the $p\pi$ invariant mass side-bands. The dashed line shows the sum of the backgrounds described in the text. The dotted line shows the Monte Carlo prediction for $\Lambda_c^+ \to \Lambda e^+ \nu_l$ normalized to the data after subtraction of the backgrounds. The solid line shows the sum of the Monte Carlo prediction and the backgrounds.

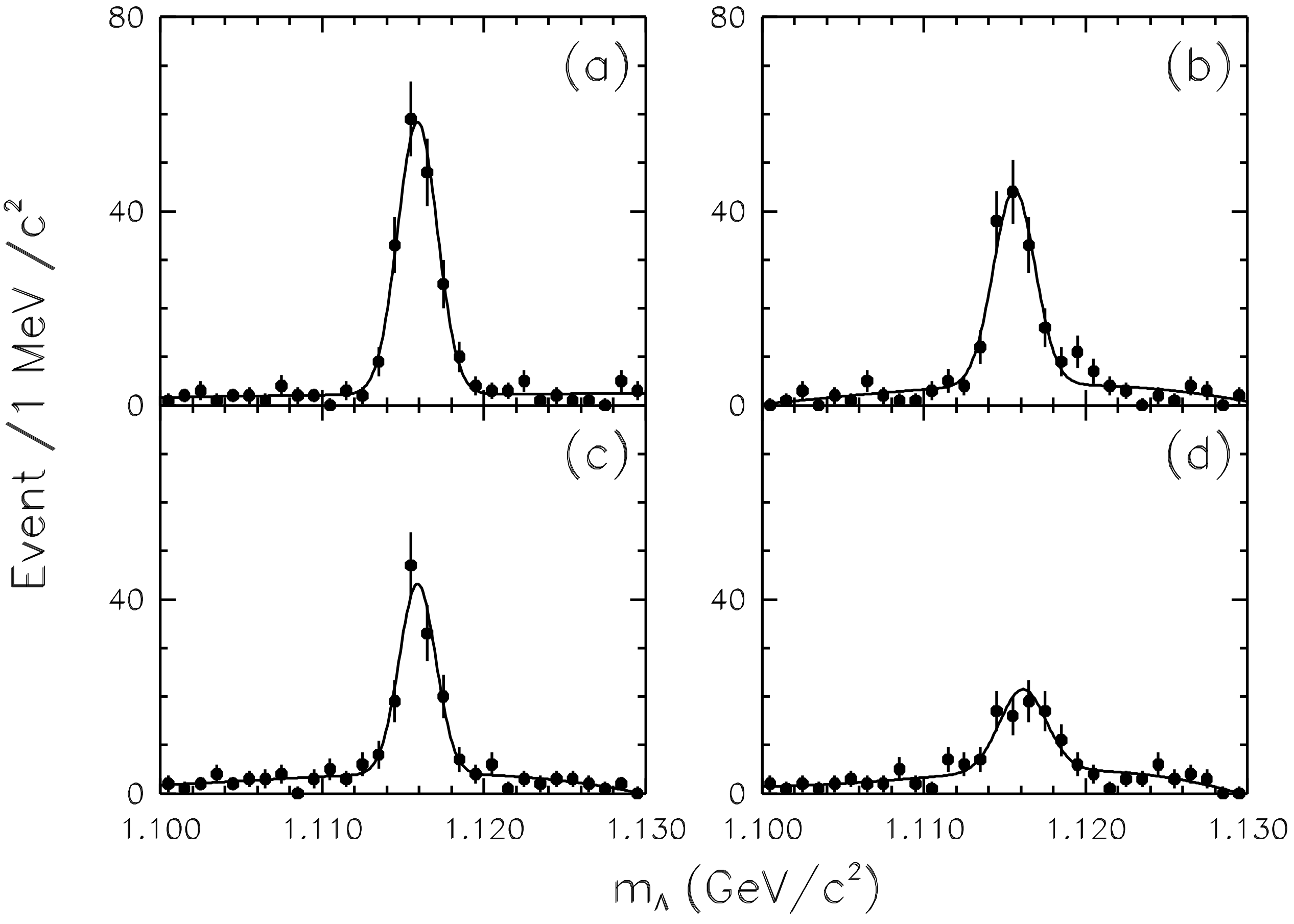

FIG. 3. The $p\pi$ invariant mass for values of $\cos\Theta_\Lambda$. (a) $-1.0 \leq \cos\Theta_\Lambda < -0.5$, (b) $-0.5 \leq \cos\Theta_\Lambda < 0.0$, (c) $0.0 \leq \cos\Theta_\Lambda < 0.5$, (d) $0.5 \leq \cos\Theta_\Lambda \leq 1.0$.

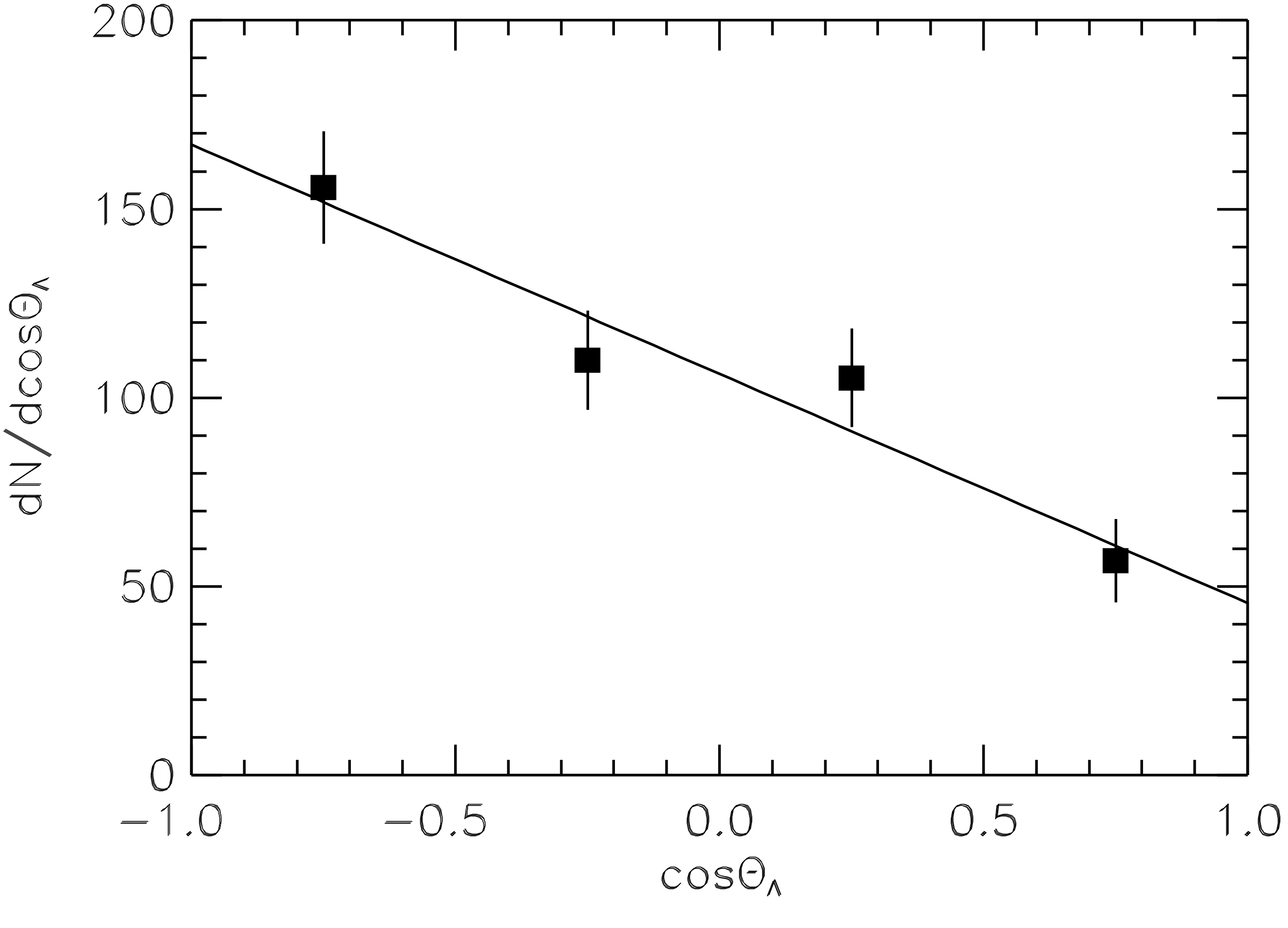

FIG. 4. $dN/d\cos\Theta_\Lambda$ with efficiency correction and background subtraction.